\documentclass[conference]{IEEEtran}
\makeatletter
\def\ps@headings{%
\def\@oddhead{\mbox{}\scriptsize\rightmark \hfil \thepage}%
\def\@evenhead{\scriptsize\thepage \hfil \leftmark\mbox{}}%
\def\@oddfoot{}%
\def\@evenfoot{}}
\makeatother
\usepackage{graphicx}
\usepackage{times}
\usepackage{graphics}
\usepackage[export]{adjustbox}
\usepackage{subfig}
\usepackage{makecell}
\usepackage{url}
\usepackage{listings}
\usepackage{fancyvrb}
\usepackage{framed}
\usepackage[listings,skins]{tcolorbox}
\usepackage{lipsum} % just for the example
\makeatletter
\newcommand{\verbatimfont}[1]{\renewcommand{\verbatim@font}{\ttfamily#1}}
\makeatother

\usepackage{epsfig}
\usepackage{caption}
\usepackage{varwidth}
\usepackage{algorithm}
\usepackage{algorithmicx,algpseudocode}

\usepackage{mwe} 
\newcommand{\CASE}[1]{\STATE \textbf{case} #1\textbf{:} \begin{ALC@g}}
\newcommand{\ENDCASE}{\end{ALC@g}}

\newcommand{\DEFAULT}{\STATE \textbf{default:} \begin{ALC@g}}
\newcommand{\ENDDEFAULT}{\end{ALC@g}}
\newcommand{\DEFAULTLINE}[1]{\STATE \textbf{default:} }
\usepackage{color}
\usepackage{multirow}
\usepackage{multicol}
\usepackage{hhline}% http://ctan.org/pkg/hhline
\usepackage{amsthm}
\usepackage{comment}
\usepackage{tabu}
\usepackage{diagbox}
\usepackage{amsmath,amsfonts,amssymb,amsthm,epsfig,epstopdf,url,array}

\theoremstyle{plain}

\theoremstyle{definition}

\theoremstyle{remark}

\IEEEoverridecommandlockouts
\begin{document}

\title{LAMP: Prompt Layer 7 Attack Mitigation With Programmable Data Planes}
\author{
Garegin Grigoryan\\ gg5996@rit.edu  \\Rochester Institute of Technology\IEEEauthorrefmark{1}\thanks{\IEEEauthorrefmark{1}This work was done when Garegin Grigoryan was a student at Clarkson University.}  \and 
Yaoqing Liu\\ liu@clarkson.edu \\
Clarkson University
}
%\date{}
\maketitle

\pagestyle{empty}

\begin{abstract}
	While there are various methods to detect application layer attacks or intrusion attempts on an individual end host, it is not efficient to provide all end hosts in the network with heavy-duty defense systems or software firewalls. In this work, we leverage a new concept of programmable data planes, to directly react on alerts raised by a victim and prevent further attacks on the whole network by blocking the attack at the network edge. We call our design LAMP, Layer 7 Attack Mitigation with Programmable data planes. We implemented LAMP using the P4 data plane programming language and evaluated its effectiveness and efficiency in the Behavioral Model (bmv2) environment.
\end{abstract}

\section{Introduction}
\label{sec:intro}
%1. Description of App layer attacks: who is the target
%2. How they are conducted
%3. Statistics, rise
%4. Why it is so bad 

Layer 7 attacks target the resources of the application layer, such as web and database servers. A victim of Layer 7 attack might be a data center, a cloud service or a private network. 
There are multiple scenarios where Layer 7 attacks can disrupt the services provided by a network. For example, a DDoS attack known as HTTP flood exhausts web servers and databases of a network by sending a great amount of POST or GET requests. The attack can be reinforced, when it is conducted by large botnets, i.e., a network of compromised devices controlled by an attacker. IoT (Internet of Things) devices are the most attractive victims to be lured into botnets, because of various security flaws. In particular, most often the users of IoT devices do not change the default login and password embedded by the manufacturer, which makes the IoT devices vulnerable to dictionary attacks~\cite{mahmoud2015internet}. Moreover, IoT devices possess limited CPU resources for identifying a malicious user or an intrusion attempt~\cite{spognardi2017analysis}. 

The number of attacks at the application layer is growing, according to the "Global DDoS Threat Landscape Report"~\cite{incapsula}. Current detection techniques for application layer attacks include wide range of measures including (1) Pattern analysis of HTTP requests; (2) Browsing behavior analysis using web logs; (3) Geo-location analysis of web clients; (4) Machine learning pre-profiling legitimate traffic; (5) Application layer challenges, such as as CAPTCHA; (6) JavaScript engine authentication and many others (\cite{infosec, ndibwile2015web, bronte2017mitigating, prabha2010mitigation, wang2017skyshield}).  All of these techniques require application data analysis with high computational capabilities, which are usually available at the powerful end hosts. Upon detection, the malicious traffic can be dropped based on their source IP addresses or other attributes, such as TCP/UDP ports or HTTP request information. Normally, there are two ways to achieve this goal with the help of a victim end host. The first one is individual or local defense, where the machine running the application installs certain firewall/IDS rules to block the attack traffic. One of the main drawbacks of this approach is that the identified attacking information cannot be re-used by other end hosts in the same network. The second approach is cooperative or global defense, where a Software Defined Networking (SDN) controller can be used to collect the detection results and to install the corresponding OpenFlow-like rules to SDN-enabled switches. However, in this case, an SDN controller introduces additional complexity and overhead to the network operations. In addition, security of the SDN controller itself is another concern.  

To enable prompt, cooperative, and efficient mitigation of Layer 7 attacks, in this work, we introduce a new approach by leveraging the Protocol Independent Switch Architecture (PISA)~\cite{bosshart2014p4}. PISA allows us to program data planes directly without involving a centralized controller by using parser engines, match-action tables, ingress and egress pipelines in P4 language~\cite{bosshart2014p4}. More specifically, we design LAMP, \underline{L}ayer 7 \underline{A}ttack \underline{M}itigation with \underline{P}rogrammable data planes. In LAMP, we track the path of each flow coming into the victim network. If an end host application detects an intrusion attempt, it generates an attack alert by embedding a signal flag and the attacker's IP address in the IP option field of the reply packet. We assume that such application has privileges to modify IP packets' fields of the alert message. The alert is sent to the closest switch directly, which eventually forwards the packet to the ingress switch that carried the original malicious traffic.  Upon receiving the alert, the ingress switch modifies its flow control policy to block the subsequent traffic from the attacker. Our new mitigation strategy yields at least three advantages: (1) It enables network-wide cooperative detection and mitigation of attacks. The detection results obtained by one end host can be re-used to benefit services for the entire network; (2) The volume of in-network malicious traffic is considerably reduced, since the network edge quickly blocks them; and (3) It enables lightweight and efficient network operations compared to existing SDN approaches, where an SDN controller is required to bridge the gap between application and network layer services. In addition, SDN incurs many additional messages, necessary to establish connections between the SDN controller and end hosts.
% (1) Once an edge switch takes measures to block the flow, every end host in a network will be defended from that flow; (2) The inner links of a network will be freed from the malicious flows; (3) LAMP minimizes the number of entries to be installed in the switches to block malicious flows, since LAMP installs those entries at the edge switches only; (4) Unlike attack mitigation techniques using SDN, LAMP is implemented entirely in the data plane of a switch which excludes the possible attack on the controller. 
%(5) Data plane code for all the switches in LAMP is homogeneous, which simplifies the maintenance of the network.
% Can be added: LAMP does not modify the inner packets of a network (75% of the traffic)
Overall, we made the following contributions in this work:

(1) We designed LAMP, a new cooperative framework for prompt and efficient mitigation of Layer 7 attacks without the involvement of a centralized SDN controller;

(2) To the best of our knowledge, it is the first time that we designed a Layer 7 intrusion mitigation solution by leveraging the new concept of programmable data planes;

(3) We implemented LAMP in P4~\cite{bosshart2014p4}, the language for programming the data plane of a switch. LAMP can re-use the same P4 code for all the switches in a network for the mitigation tasks. 
%We give the detailed description of the P4 code in this paper;

(4) We emulated LAMP in the \textit{bmv2} model~\cite{bmv2}, a virtual environment designed for testing P4-programmed switches. Our comparison with a similar SDN architecture shows that LAMP bears much less operational complexity and minimizes the number of malicious application messages reaching end hosts once an attack is detected.

%The rest of this paper is organized as follows: ... 

%In this work, we propose a c

%The application layer (Layer 7 of the OSI model) attacks mostly target web and cloud services, as well as Layer 7 protocols such as HTTP, HTTPS, DNS and SSH []. Such attacks are able to bypass network's defense infrastructure since they mimic a legitimate traffic's behavior on a network layer. Moreover, identifying malicious flows requires a deep analysis of packets' payload by Firewalls, IDSes (Intrusion Defense Systems) or IPSes (Intrusion Prevent Systems).
\section{Related Work}
\label{sec:related}
In ~\cite{norman2017security}, Norman $et$ $al.$ studied application layer protocols used in modern IoT devices and their vulnerabilities. De Donno $et$ $al.$ in~\cite{spognardi2017analysis} presented a deep survey on DDoS attacks against the IoT. Particularly, the authors analyzed Mirai, an attack that was conducted using large number of IoT devices, compromised with application layer dictionary attacks and lured into the botnets. In~\cite{ndibwile2015web, prabha2010mitigation, wang2017skyshield,  bronte2017mitigating}, the authors proposed various Layer 7 attack detection techniques, such as additional client tests (CAPTCHA, passwords, puzzles, JavaScript authentication); web logs analysis and building a profile of a legitimate user; analysis of the visiting history of clients; traffic classification. LAMP is compatible with all of these approaches since it is designed for fast mitigation of attacks after they are detected by a member(s) of the network. 

Giotis $et$ $al.$ in~\cite{giotis2014combining} presented an SDN architecture for flow-based anomaly detection and installation of the mitigation rules on the edge switches. The mitigation rules are dropping the packets based on their source and destination IP addresses. Lim $et$ $al.$~\cite{lim2014sdn} design DDoS blocking solution that changes the IP address of the victim and redirects the legitimate connections, in order to mitigate large botnet attacks. In our work, we present LAMP architecture, that blocks malicious traffic at the edge switches using programmable data planes.
\section{Design}
\begin{figure*}[t!]
	\centering
	\captionsetup{justification=centering, width=.28\linewidth}
	\subfloat[Initial state]
	\centering
	\label{fig:scenario1}
	\includegraphics[width=.24\linewidth]{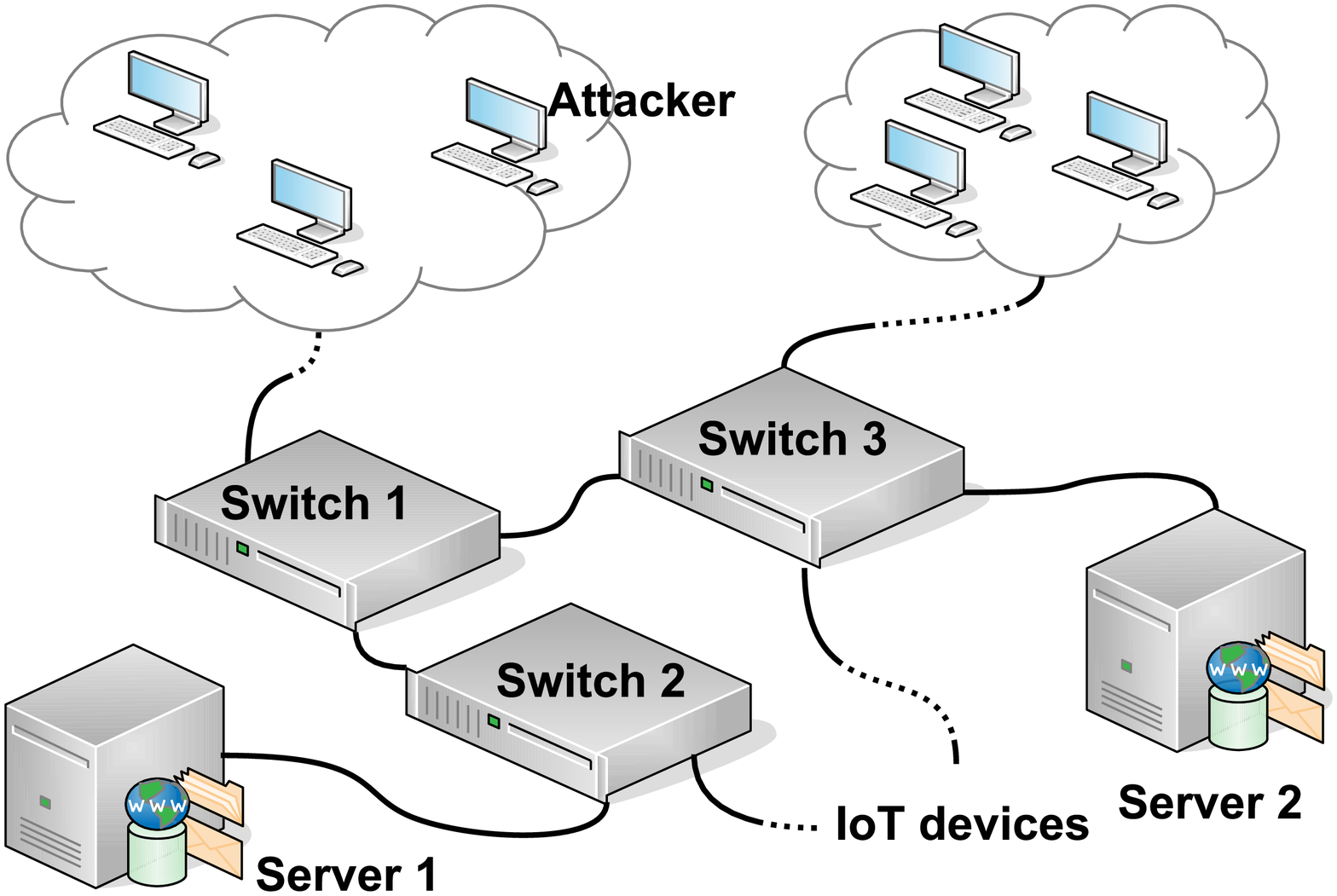}
	%\caption*{{\small }Round one \tab Round two steps} 
	\captionsetup{justification=centering, width=.21\linewidth}
	\subfloat[Scanning the hosts of the victim network by \textit{Attacker}]
	\centering
	\includegraphics[width=.24\linewidth]{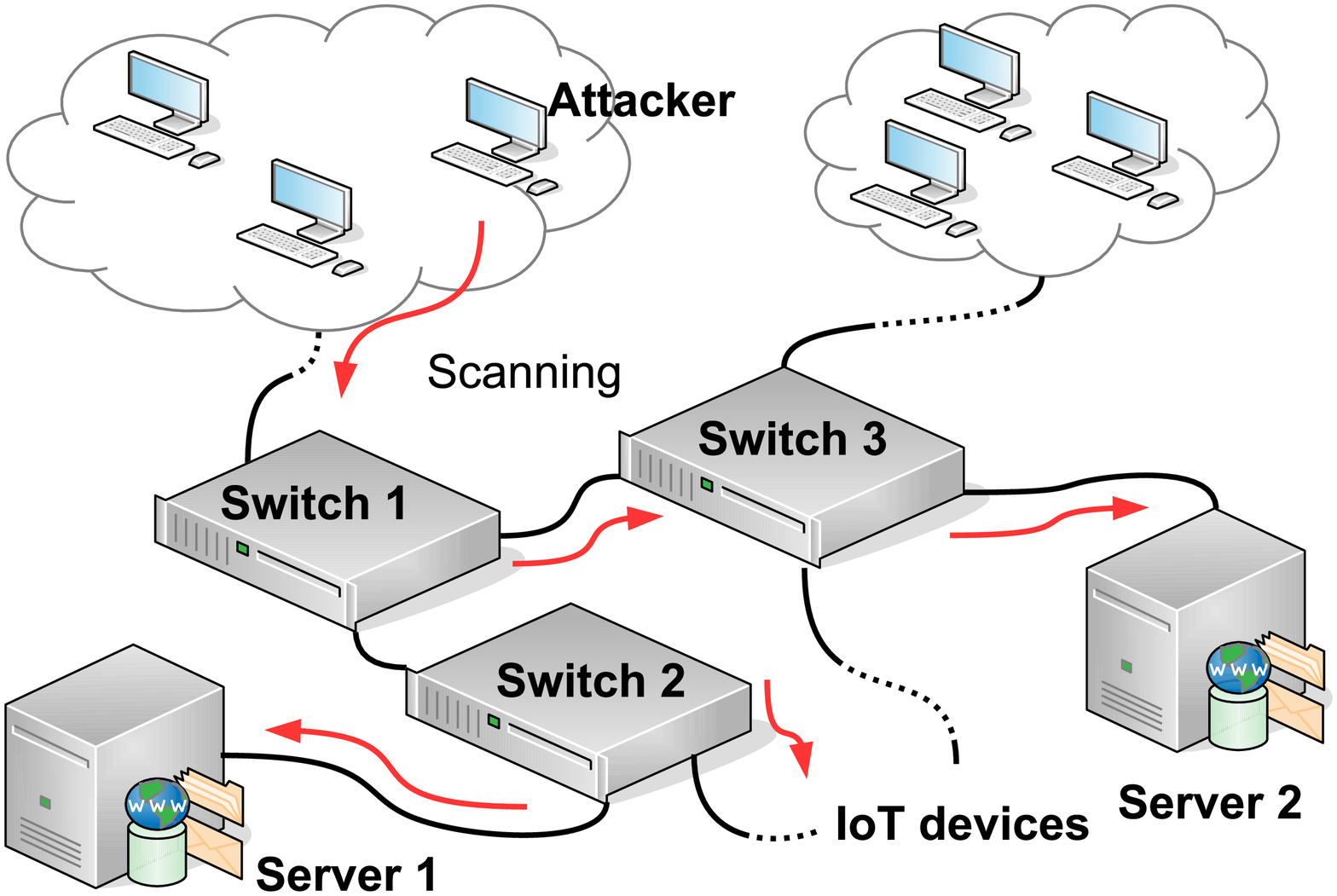}
	%\caption*{{\small }Round one \tab Round two steps} 
	\label{fig:scenario2}
	\subfloat[\textit{Server 2} detects the scanning attempt and sends the alert]
	\centering
	\includegraphics[width=.25\linewidth]{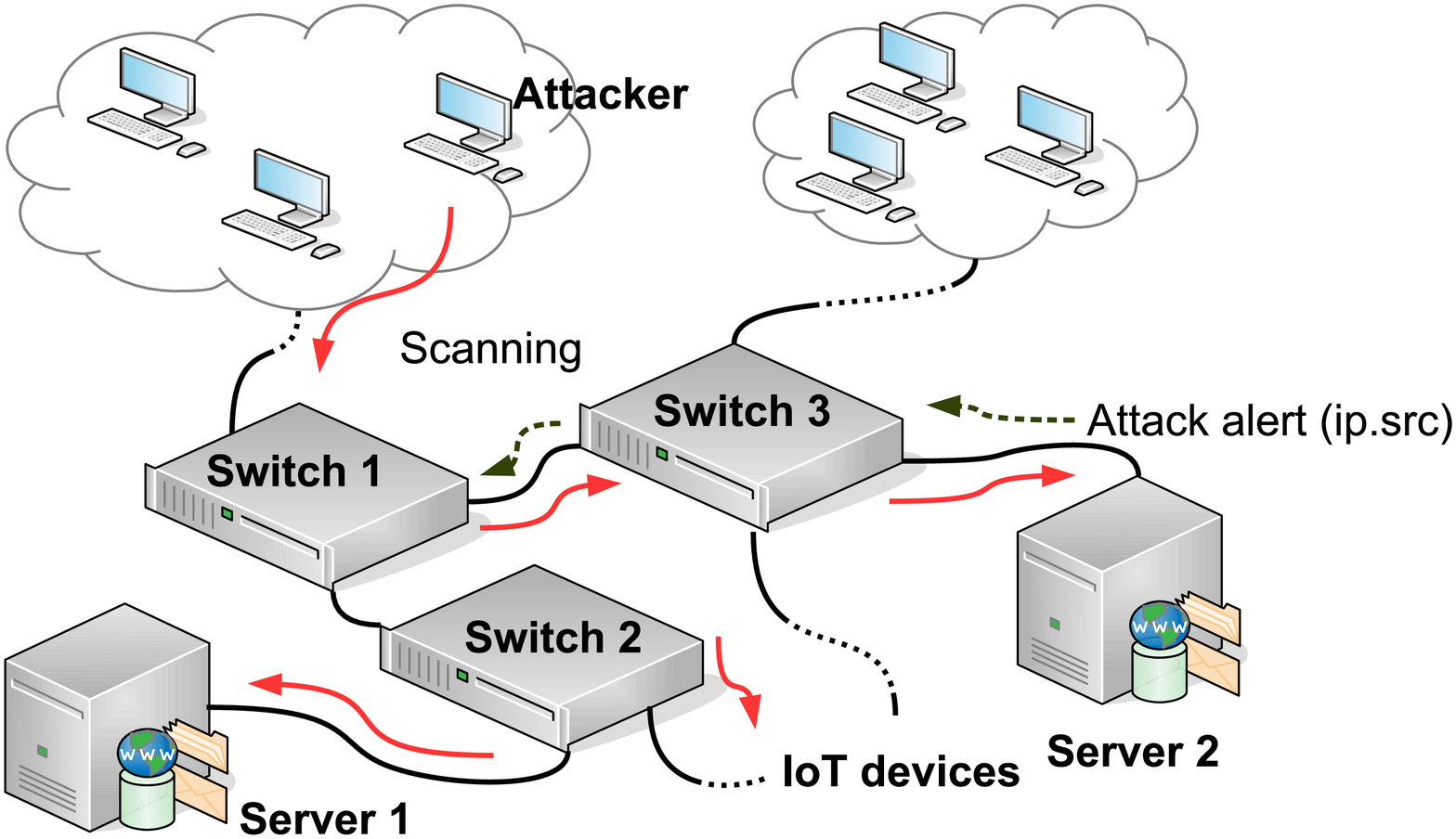}
	%\caption*{{\small }Round one \tab Round two steps} 
	\label{fig:scenario3}
	\subfloat[Attack is blocked at the edge \textit{Switch 1}]
	\centering
	\includegraphics[width=.24\linewidth]{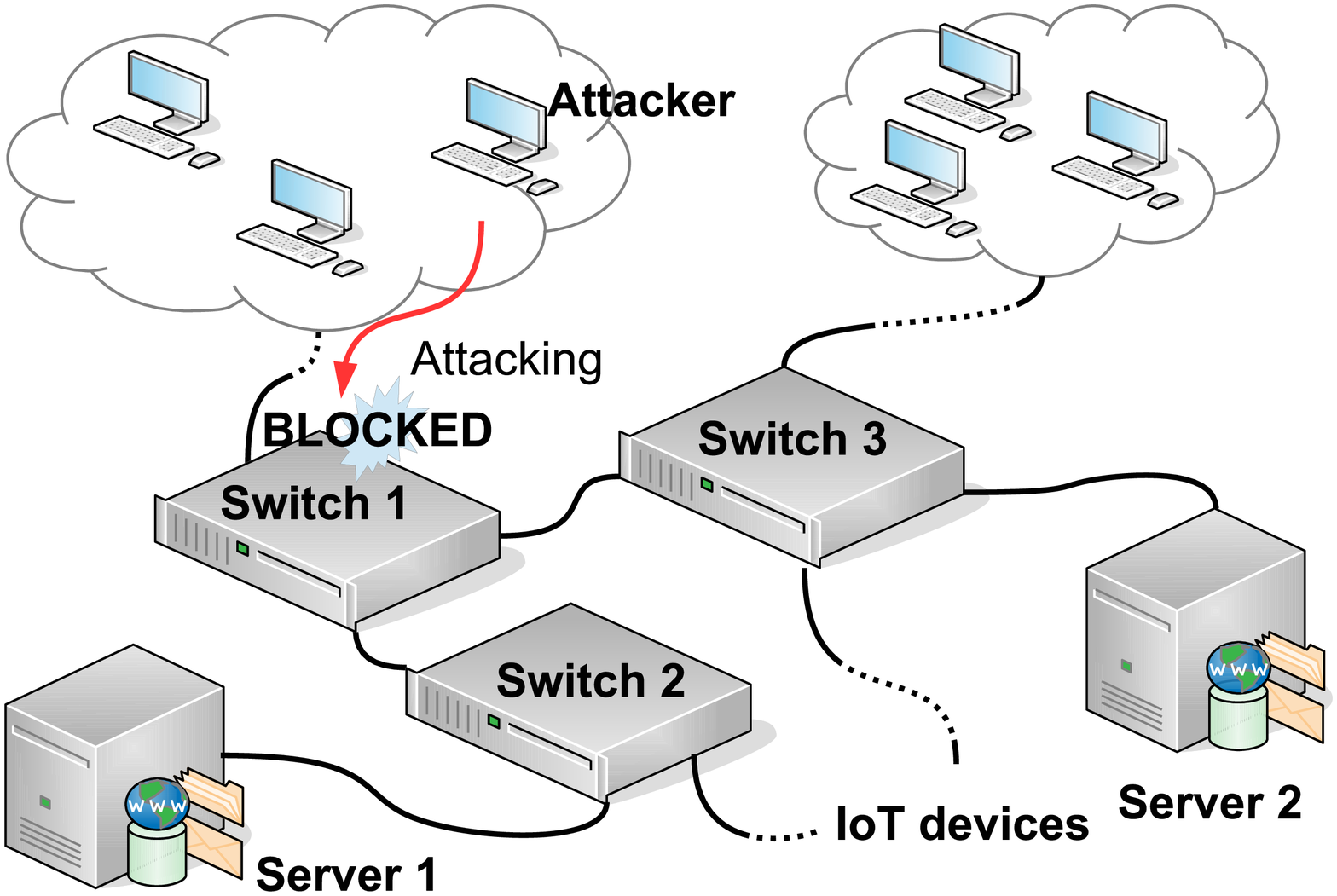}
	%\caption*{{\small }Round one \tab Round two steps} 
	\label{fig:scenario4}
	\captionsetup{justification=centering, width=.9\linewidth}
	\caption{Scenario of Layer 7 attack mitigation with LAMP}
	\label{fig:scenario}
	\vspace{-5mm}
\end{figure*}
We demonstrate an example of Layer 7 attack mitigation with LAMP on Figure~\ref{fig:scenario}. The attacker starts scanning resources of the victim network in Figure~\ref{fig:scenario}(b), e.g., a random dictionary attack in order to login into the most vulnerable end hosts. Since the attack is conducted against the application layer, the network layer devices, such as switches and routers, are unable to detect it, and they allow the packets to reach the end hosts. To track the entrance point of each external flow, \textit{Switch 1} encapsulates its ID into the option field of the incoming IP packet's headers. The option is given a special type \textit{INGRESS\_SWITCH\_INFO} to differentiate it from other possible options. Later, the switch ID and its option are dumped as the last switch on the path forwards the packet to the final destination. Meanwhile, that switch records the mapping between the flow's source IP address and the edge (ingress) switch's ID.

In our scenario, the end-host \textit{Server 2} detects the scanning attempt and sends the attack alert message back into the network (Figure~\ref{fig:scenario}(c)). The alert is encapsulated inside \textit{ATTACK\_ALERT} option. The attack alert contains the IP source address of the attacker. The switch (\textit{Switch 3}) that receives the alert finds the corresponding ingress switch (\textit{Switch 1}), adds the switch ID to the IP option, changes its type to \textit{FORWARD} and sends the packet to the next hop towards that switch (\textit{Switch 1}). Lastly, \textit{Switch 1} installs necessary entries to drop the packets from the attacker before they enter the network. To implement LAMP using the P4 language, we need to program the following components over a programmable data plane: (1) The parser; (2) Match-action tables; (3) Ingress and egress flows of a switch. In the rest of this section, we give the detailed description of our modifications in each of those components.

\subsection{Parser}
\begin{figure}
	\begin{center}
		\includegraphics[width=1\linewidth]{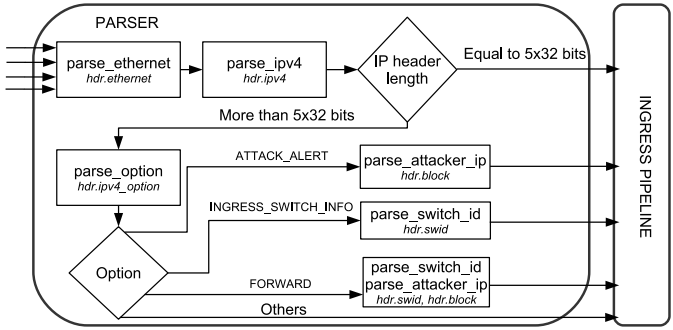}
		\caption{\label{fig:parser}Parser in LAMP}
	\end{center}
	\vspace{-11mm}
\end{figure}
To implement the parser, we first define (1) The headers that should be decoupled and read from each packet; (2) The possible option numbers used in LAMP. We create 3 types of options with the following option numbers from the unassigned range: \textit{ATTACK\_ALERT} = 31; \textit{INGRESS\_SWITCH\_INFO} = 29;  \textit{FORWARD} = 27. 

In addition to Ethernet (\textit{ethernet\_t}), IP (\textit{ipv4\_t}) and the standard IP option header (\textit{ipv4\_option\_t}), we defined the switch header (\textit{switch\_t}) and the alert header (\textit{block\_t}), that shall contain the edge switch's ID and attacker's IP address respectively. \textit{switch\_t} and \textit{block\_t} are placed within the payload of \textit{ipv4\_option\_t}, so LAMP's parser does not need to decouple the Layer 4 (the Transport Layer) headers. In the meantime, for the internal packets of a network, the parser will not decouple \textit{switch\_t} or \textit{block\_t} headers, because those packets shall not contain any of the above-mentioned IP options. Thus, we exclude the additional overhead for the internal traffic, which in some cases constitutes more than 75\% of the total traffic within a network~\cite{itjungle}.
%this can be removed
LAMP's parser's state transitions are illustrated on Figure~\ref{fig:parser}. %We present the P4 code of the LAMP's parser in Listing~\ref{code2}. %Note, than in case of the \textit{FORWARD} option, parser reads both switch ID and attacker's IP, even if a switch is not the final receiver of the request. This could be avoided if we matched the \textit{switchID.swid} of the option against the ID of the switch that parses that option.  However, that would require us to hardcode switch's ID into a constant variable, which would contravene with our desire to design a homogeneous data plane code for the whole network. Therefore, on receiving  \textit{FORWARD}, each switch in LAMP parses both headers (\textit{hdr.switchID} and \textit{hdr.block}), although it will use  \textit{hdr.block.attackerIP} if only its switch ID, defined by control plane, is equal to \textit{switchID.swid}.

\subsection{Match-action tables}
\begin{comment}
\lstset{language=C,frame=single}
\begin{lstlisting}[tabsize=2, escapeinside={(*}{*)}, basicstyle=\footnotesize,float,caption=Tables in LAMP,label=code3,belowskip=-2\baselineskip]
/* ipv4_lpm table (see [15]) */
.............
/* swid_add, swid_remove, swid_forward */
table swid_add {
key = {
standard_metadata.ingress_port: exact;
}
actions = {
add_swid; NoAction;
}
size = 16;
default_action = NoAction();
}
table swid_remove{
key = {
standard_metadata.egress_spec: exact;
ipv4_option.option: exact;
}
actions = {
remove_swid; NoAction;
}
size = 16;
default_action = NoAction();
}

table swid_forward {
key = {
hdr.switchID.swid: exact;
}
actions = {
block; ipv4_forward; NoAction;
}
size = 16;
default_action = NoAction();
}
\end{lstlisting}
\end{comment}

In LAMP, match-action tables are used for (1) Forwarding the packets based on their IP destination address; (2) Adding the edge switch's ID to a packet that comes from outside the network; (3) Removing the edge switch's ID from a packet whose next hop is an end host; (4) Forwarding the attack alert based on the edge switch's ID. While the tables' structure and actions are defined in the data plane, the content of the tables is regulated by the control plane. We assume that the centralized or a distributed control plane can automatically populate match-action tables using API generated by P4 compiler. %Listing~\ref{code3} shows the tables written in P4 and we give the detailed description of each table as follows.

(1) \textit{ipv4\_lpm}: As in the "Simple Switch" presented by P4 designers~\cite{ipv4_forward}, table \textit{ipv4\_lpm} contains the prefixes and the corresponding next hops (output port and MAC address information). The destination IP address of a packet will be matched against one of those prefixes in order to get the next hop. Based on the match, the packet is either forwarded (action \textit{ipv4\_forward}) or dropped. 

(2) \textit{swid\_add}: Contains the mapping of the current switch's ID and the list of ports that connect the switch to an external network. On receiving a packet from one of those ports, the switch will attach its ID to the packet (action \textit{add\_swid}). In addition, the packet will be marked as "to be checked" using the packet's metadata field (\textit{meta.check\_source\_ip}=1). The table is empty for internal switches.

% host_switch_port
% network_facing

(3) \textit{swid\_remove}: Contains ports that connect to end hosts and the option numbers we specified for Layer 7 attack mitigation. LAMP needs to remove our previously attached \textit{INGRESS\_SWITCH\_INFO} option via \textit{remove\_swid} action from the packets that are reaching an end host at their next hop. Meanwhile, the switch needs to store the mapping between the attached ingress switch ID and the hashed value of the packet's source IP address. The table should be empty for switches that are not directly connected to end hosts. 

(4) \textit{swid\_forward}: Contains the list of destination switch IDs, the next hop ports and the corresponding MAC addresses in the network. This table may incur two different actions based on the value of packet's destination switch ID \textit{hdr.switchID.swid}: (a) If the value equals to the current switch ID, run the action \textit{block}, that will install a drop entry into the blacklist hash table for the attached source IP address in the alert message; (b) Otherwise, run the action \textit{ipv4\_forward}, that will forward the packet to the next hop towards the destination switch whose ID equals to \textit{hdr.switchID.swid}.  To enable such workflow, the control plane needs to correctly initialize \textit{swid\_forward} table.

\subsection{Control flow}
\begin{figure}
	\begin{center}
		\includegraphics[width=1\linewidth]{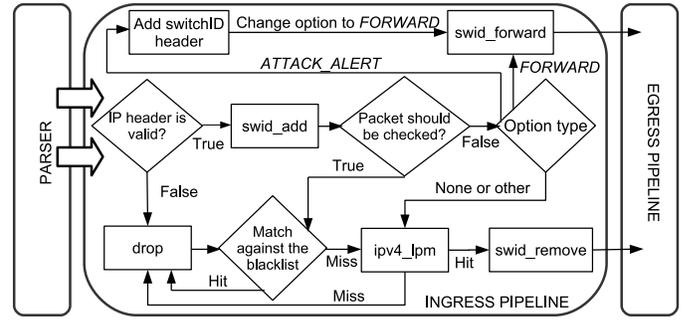}
		\caption{\label{fig:ingress}Ingress pipeline in LAMP}
	\end{center}
	\vspace{-7mm}
\end{figure}
\begin{figure}
	\begin{center}
		\includegraphics[width=1\linewidth]{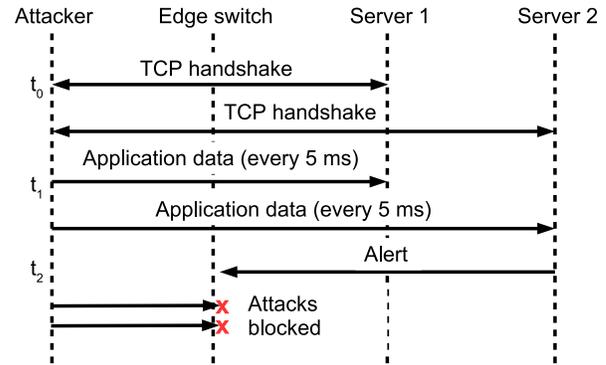}
		\captionsetup{justification=centering}
		\caption{\label{fig:scenario_eval}Evaluation scenario}
	\end{center}
	\vspace{-9mm}
\end{figure}

\subsubsection{Ingress flow} After a packet is parsed (Figure~\ref{fig:parser}), its header fields are transmitted to the ingress pipeline as shown in Figure~\ref{fig:ingress}. At first, LAMP checks the validity of the IP header. If valid, the packet's input port is matched against \textit{swid\_add} table. Two cases may happen: (1) If the port faces an external network, the current switch ID is attached to the header option of the packet. In addition, the packet's source IP address will be matched against the blacklist. If a match is found, it indicates that this packet was from an attacker and LAMP drops it; (2) Otherwise, the packet's IP option is checked. If there is an \textit{ATTACK\_ALERT} option, the switch first finds the corresponding ingress switch and adds its ID to the packet header option. Then it changes the option to \textit{FORWARD} and forwards the packet towards that ingress switch.
% and then matches the packet against \textit{swid\_forward} match-action table. 
%The next is to check \textit{swid\_forward} match-action table and the \textit{FORWARD} option, if both , then LAMP forwards the packet to the next switch.
%the switch matches the packet against \textit{swid\_forward} table without header modification. 
In case the packet does not have above-mentioned  options, it is matched against \textit{ipv4\_lpm} table to obtain the next hop port and MAC address. Lastly, the packet is matched against \textit{swid\_remove} table to remove option \textit{INGRESS\_SWITCH\_INFO} if it exists and the next hop for the packet is an end host.

\subsubsection{Egress flow}
For the egress flow, LAMP simply de-parses the packet header in the following order:  \textit{hdr.ethernet}, \textit{hdr.ipv4}, \textit{hdr.ipv4\_option}, \textit{hdr.block}, \textit{hdr.switchID}. P4 program automatically checks if each of these headers is valid and omits the header if not. We omit P4 code listings ingress control flow and the deparser due to the space constraints.

%because of the limited space. However, it can be viewed online [cite].
% to implement the abovementioned workflow, we needed to modify parse mashine, match-action tables
%(i.e., \textit{hdr.$<$header\_name$>$.isValid()} returns \textit{True}) 

\section{Evaluation}
\label{sec:evaluation}
%where we simulated
%topology
%reaction time
We emulated LAMP in the Behavioral Model (bmv2)~\cite{bmv2}, which provides a P4 software switch with the compiler using Mininet virtual environment~\cite{kaur2014mininet}. We compared LAMP with a similar architecture, implemented using SDN and OpenFlow~\cite{mckeown2008openflow} in Mininet. The topology for the experiment is similar to one illustrated on Figure~\ref{fig:scenario}.  We used the following scenario for the experiment (see Figure~\ref{fig:scenario_eval}): At the moment $t_0$, the \textit{Attacker} establishes TCP connection with \textit{Server 1} and \textit{Server 2}; at $t_{1}$, it starts sending packets with invalid HTTP requests with a rate 200 packets/s. As soon as \textit{Server 2} detects the attack, it sends an alert message into the network ($t_2$). In case of LAMP, the message reaches the edge switch, where a blocking entry is installed into the blacklist. In case of SDN, the message is captured by an SDN controller, which figures out the corresponding ingress switch and installs a drop rule into its OpenFlow table. %Depending on how fast these rules are be installed, the vulnerable \textit{Victim 2} receives less packets from \textit{Attacker}.
In our experiment, we emulated 30 attacks for both LAMP and SDN architectures. 
%In each attack, We programmed the attacker to flood messages to both victims/servers shown in Figure~\ref{fig:topology}. We assume that \textit{victim 1} is able to detect the attack and send alert messages to the network. 
\begin{table}
	\centering
	\begin{tabular}{| l | l | l |}
		\hline
		\diagbox{Measurement}{Architecture}  & LAMP  & SDN \\ \hline
		Total	&	278 & 1288    \\ \hline
		Maximum & 10 & 106    \\ \hline
		Minimum	&	5 & 5   \\ \hline
		Average	&	9 & 43    \\ \hline
	\end{tabular}
	\captionsetup{justification=centering}
	\caption{The number of invalid HTTP requests that\\ \textit{Server 1} received in SDN and LAMP architectures}
	\label{tab:results}
	\vspace{-8mm}
\end{table}
We intended to observe how many messages can go through the network to reach the other victim before the attack can be blocked. Interestingly, unlike LAMP, the SDN emulation produced fluctuating results, as it can be seen in Table~\ref{tab:results}.  Overall, in LAMP, \textit{Server 1} received only 278 invalid HTTP requests, 1010 packets less than that received by \textit{Server 1} in the SDN architecture. Moreover, the maximum number of such requests during a single attack in LAMP is 10, while in SDN it reached 106. But, in some cases, both LAMP and SDN controller acted fast enough to block \textit{Attacker}, so \textit{Server 1} received only 5 invalid HTTP requests. On average, in SDN, \textit{Server 1} received 80\% more invalid HTTP requests than that in LAMP. We attribute it to the use of a centralized controller that introduces additional overhead and complexity. In LAMP, attack alerts are processed fully in the data plane, which makes attack mitigation significantly faster. In the meantime, other factors might have affected the performance of two emulated architectures, such as the differences and deficiencies of P4 and SDN implementations in Mininet.

\section{Conclusion}
\label{sec:conclusion}
In this work, we presented LAMP, an architecture for Layer 7 attack mitigation with programmable data planes. To the best of our knowledge, for the first time we leveraged Protocol Independent Switch Architecture (PISA) to design a cooperative mitigation solution against the application layer attacks. We presented the detailed solution of the new mitigation architecture in LAMP, including the modified parser, match-action tables and the ingress flow. We implemented LAMP in the P4 language and emulated it in the Mininet virtual environment. Compared to a similar Software Defined Networking architecture, LAMP mitigates the Layer 7 attacks more quickly and minimizes the number of malicious application layer messages that are sent to victims of the same network. 
\vspace{-2.3mm}
\bibliographystyle{IEEEtran}
\bibliography{net}

% Generated by IEEEtran.bst, version: 1.12 (2007/01/11)
\begin{thebibliography}{10}
\providecommand{\url}[1]{#1}
\csname url@samestyle\endcsname
\providecommand{\newblock}{\relax}
\providecommand{\bibinfo}[2]{#2}
\providecommand{\BIBentrySTDinterwordspacing}{\spaceskip=0pt\relax}
\providecommand{\BIBentryALTinterwordstretchfactor}{4}
\providecommand{\BIBentryALTinterwordspacing}{\spaceskip=\fontdimen2\font plus
\BIBentryALTinterwordstretchfactor\fontdimen3\font minus
  \fontdimen4\font\relax}
\providecommand{\BIBforeignlanguage}[2]{{%
\expandafter\ifx\csname l@#1\endcsname\relax
\typeout{** WARNING: IEEEtran.bst: No hyphenation pattern has been}%
\typeout{** loaded for the language `#1'. Using the pattern for}%
\typeout{** the default language instead.}%
\else
\language=\csname l@#1\endcsname
\fi
#2}}
\providecommand{\BIBdecl}{\relax}
\BIBdecl

\bibitem{mahmoud2015internet}
R.~Mahmoud, T.~Yousuf, F.~Aloul, and I.~Zualkernan, ``{Internet of things (iot)
  security: Current status, challenges and prospective measures},'' in
  \emph{Internet Technology and Secured Transactions (ICITST), 2015 10th
  International Conference for}.\hskip 1em plus 0.5em minus 0.4em\relax IEEE,
  2015, pp. 336--341.

\bibitem{spognardi2017analysis}
A.~Spognardi, M.~De~Donno, N.~Dragoni, and A.~Giaretta, ``Analysis of
  ddos-capable iot malwares,'' \emph{Annals of Computer Science and Information
  Systems}, vol.~11, pp. 807--816, 2017.

\bibitem{incapsula}
\BIBentryALTinterwordspacing
{Imperva Incapsula}, ``{Q1 2017 Global DDoS Threat Landscape Report},'' 2017.
  [Online]. Available:
  \url{https://www.incapsula.com/blog/q1-2017-global-ddos-threat-landscape-report.html/}
\BIBentrySTDinterwordspacing

\bibitem{infosec}
\BIBentryALTinterwordspacing
{Infosec Institute}, ``{Layer 7 DDoS Attacks: Detection And Mitigation},''
  2013. [Online]. Available:
  \url{http://resources.infosecinstitute.com/layer-7-ddos-attacks-detection-mitigation/}
\BIBentrySTDinterwordspacing

\bibitem{ndibwile2015web}
J.~D. Ndibwile, A.~Govardhan, K.~Okada, and Y.~Kadobayashi, ``Web server
  protection against application layer ddos attacks using machine learning and
  traffic authentication,'' in \emph{Computer Software and Applications
  Conference (COMPSAC), 2015 IEEE 39th Annual}, vol.~3.\hskip 1em plus 0.5em
  minus 0.4em\relax IEEE, 2015, pp. 261--267.

\bibitem{bronte2017mitigating}
R.~Bronte, H.~Shahriar, and H.~M. Haddad, ``Mitigating distributed denial of
  service attacks at the application layer,'' in \emph{Proceedings of the
  Symposium on Applied Computing}.\hskip 1em plus 0.5em minus 0.4em\relax ACM,
  2017, pp. 693--696.

\bibitem{prabha2010mitigation}
S.~Prabha and R.~Anitha, ``Mitigation of application traffic ddos attacks with
  trust and am based hmm models,'' \emph{International Journal of Computer
  Applications IJCA}, vol.~6, no.~9, pp. 26--34, 2010.

\bibitem{wang2017skyshield}
C.~Wang, T.~N. Miu, X.~Luo, and J.~Wang, ``Skyshield: A sketch-based defense
  system against application layer ddos attacks,'' \emph{IEEE Transactions on
  Information Forensics and Security}, 2017.

\bibitem{bosshart2014p4}
P.~Bosshart, D.~Daly, G.~Gibb, M.~Izzard, N.~McKeown, J.~Rexford,
  C.~Schlesinger, D.~Talayco, A.~Vahdat, G.~Varghese \emph{et~al.}, ``P4:
  Programming protocol-independent packet processors,'' \emph{ACM SIGCOMM
  Computer Communication Review}, vol.~44, no.~3, pp. 87--95, 2014.

\bibitem{bmv2}
\BIBentryALTinterwordspacing
``{Behavioral model repository},'' 2017. [Online]. Available:
  \url{https://github.com/p4lang/behavioral-model/}
\BIBentrySTDinterwordspacing

\bibitem{itjungle}
\BIBentryALTinterwordspacing
{Timothy Prickett Morgan}, ``{Cisco: Data Center Traffic To Quadruple Thanks To
  Clouds},'' 2012. [Online]. Available:
  \url{https://www.itjungle.com/2012/10/29/tfh102912-story06/}
\BIBentrySTDinterwordspacing

\bibitem{ipv4_forward}
\BIBentryALTinterwordspacing
{Barefoot}, ``{Simple switch},'' 2012. [Online]. Available:
  \url{https://github.com/p4lang/tutorials/blob/master/P4D2_2017_Spring/exercises/ipv4_forward/solution/ipv4_forward.p4}
\BIBentrySTDinterwordspacing

\bibitem{kaur2014mininet}
K.~Kaur, J.~Singh, and N.~S. Ghumman, ``Mininet as software defined networking
  testing platform,'' in \emph{International Conference on Communication,
  Computing \& Systems (ICCCS)}, 2014, pp. 139--42.

\bibitem{mckeown2008openflow}
N.~McKeown, T.~Anderson, H.~Balakrishnan, G.~Parulkar, L.~Peterson, J.~Rexford,
  S.~Shenker, and J.~Turner, ``Openflow: enabling innovation in campus
  networks,'' \emph{ACM SIGCOMM Computer Communication Review}, vol.~38, no.~2,
  pp. 69--74, 2008.

\bibitem{norman2017security}
J.~Norman and P.~Joseph, ``{Security in Application Layer Protocols of IoT:
  Threats and Attacks},'' in \emph{Security Breaches and Threat Prevention in
  the Internet of Things}.\hskip 1em plus 0.5em minus 0.4em\relax IGI Global,
  2017, pp. 76--95.

\bibitem{giotis2014combining}
K.~Giotis, C.~Argyropoulos, G.~Androulidakis, D.~Kalogeras, and V.~Maglaris,
  ``Combining openflow and sflow for an effective and scalable anomaly
  detection and mitigation mechanism on sdn environments,'' \emph{Computer
  Networks}, vol.~62, pp. 122--136, 2014.

\bibitem{lim2014sdn}
S.~Lim, J.~Ha, H.~Kim, Y.~Kim, and S.~Yang, ``{A SDN-oriented DDoS blocking
  scheme for botnet-based attacks},'' in \emph{Ubiquitous and Future Networks
  (ICUFN), 2014 Sixth International Conf on}.\hskip 1em plus 0.5em minus
  0.4em\relax IEEE, 2014, pp. 63--68.

\end{thebibliography}


% Generated by IEEEtran.bst, version: 1.12 (2007/01/11)
\begin{thebibliography}{1}
\providecommand{\url}[1]{#1}
\csname url@samestyle\endcsname
\providecommand{\newblock}{\relax}
\providecommand{\bibinfo}[2]{#2}
\providecommand{\BIBentrySTDinterwordspacing}{\spaceskip=0pt\relax}
\providecommand{\BIBentryALTinterwordstretchfactor}{4}
\providecommand{\BIBentryALTinterwordspacing}{\spaceskip=\fontdimen2\font plus
\BIBentryALTinterwordstretchfactor\fontdimen3\font minus
  \fontdimen4\font\relax}
\providecommand{\BIBforeignlanguage}[2]{{%
\expandafter\ifx\csname l@#1\endcsname\relax
\typeout{** WARNING: IEEEtran.bst: No hyphenation pattern has been}%
\typeout{** loaded for the language `#1'. Using the pattern for}%
\typeout{** the default language instead.}%
\else
\language=\csname l@#1\endcsname
\fi
#2}}
\providecommand{\BIBdecl}{\relax}
\BIBdecl

\bibitem{routeviews}
\BIBentryALTinterwordspacing
{Advanced Network Technology Center and University of Oregon}, ``The
  {RouteViews} project.'' [Online]. Available: \url{http://www.routeviews.org/}
\BIBentrySTDinterwordspacing

\bibitem{tariq2011taco}
A.~Tariq, S.~Jawad, and Z.~A. Uzmi, ``{TaCo}: Semantic {E}quivalence of {IP}
  {P}refix {T}ables,'' in \emph{Computer Communications and Networks (ICCCN),
  2011 Proceedings of 20th International Conference on}.\hskip 1em plus 0.5em
  minus 0.4em\relax IEEE, 2011, pp. 1--6.

\end{thebibliography}
\end{document}